\documentclass[preprint,preprintnumbers,amsmath,amssymb]{revtex4}

\usepackage{graphicx}
\usepackage{dcolumn}
\usepackage{bm}

\begin{document}


\title{Influence of a dynamical gluon mass in \\
the $pp$ and $\bar{p}p$ forward scattering}

\author{E.G.S. Luna,$^{1,2}$ A.F. Martini,$^{2}$ M.J. Menon,$^{2}$ A. Mihara,$^{3}$ and
A.A. Natale$\, ^{1}$}
\affiliation{
$^{1}$Instituto de F\'{\i}sica Te\'orica, Universidade Estadual Paulista, 01405-900,
S\~ao Paulo, SP, Brazil \\
$^{2}$Instituto de F\'{\i}sica Gleb Wataghin, Universidade Estadual de Campinas,
13083-970, Campinas, SP, Brazil \\
$^{3}$Instituto de F\'{\i}sica de S\~ao Carlos, Universidade de S\~ao Paulo, 13560-970,
S\~ao Carlos, SP, Brazil}


\begin{abstract}
We compute the tree level cross section for gluon-gluon elastic scattering taking into account a dynamical
gluon mass, and show that this mass scale is a natural regulator for this subprocess cross section. Using
an eikonal approach in order to examine the relationship between this gluon-gluon scattering and the elastic
$pp$ and $\bar{p}p$ channels, we found that the dynamical gluon mass is of the same order of magnitude as
the {\it ad hoc} infrared mass scale $m_{0}$ underlying eikonalized QCD-inspired models. We argue that this
correspondence is not an accidental result, and that this dynamical scale indeed represents the onset
of nonperturbative contributions to the elastic hadron-hadron scattering. 
We apply the eikonal model with a dynamical infrared mass scale to obtain predictions
for $\sigma_{tot}^{pp,\bar{p}p}$, $\rho^{pp,\bar{p}p}$, slope $B^{pp,\bar{p}p}$, and
differential elastic scattering cross section $d\sigma^{\bar{p}p}/dt$ at Tevatron and CERN-LHC energies.
\end{abstract}


\maketitle

\section{Introduction}

The increase of hadron-hadron total cross sections was theoretically predicted many years ago \cite{cheng}
and this prediction has been accurately verified by experiment \cite{hagiwara}. At present the main
theoretical approaches to explain this behavior are the Regge pole model and the QCD-inspired models.

In the Regge pole model the increase of the total cross section is attributed to the exchange of a colorless
state having the quantum numbers of the vacuum: the Pomeron \cite{pomeron}. In the QCD framework the
Pomeron can be understood
as the exchange of at least two gluons in a color singlet state \cite{low}. A simple and interesting model
for the Pomeron has been put forward where it is evidenced the importance of the QCD nonperturbative
vacuum \cite{landshoff}. One of the aspects of this nonperturbative physics appears in an infrared (IR) gluon
mass scale which regulates the divergent behavior of the Pomeron exchange.

In the QCD-inspired (or ``mini-jet") models the increase of the total cross sections is associated
with semihard scatterings of partons in the hadrons. The energy dependence of the cross sections is
driven especially by gluon-gluon scattering processes, where the behavior of the gluon
distribution function at small $x$ exhibits the power law $g(x,Q^{2}) \sim x^{-J}$ (see
\cite{block,block2,block3} and references therein). In
this case it is the gluon-gluon subprocess cross section that is potentially divergent at small transferred
momenta. The procedure to regulate this behavior is the introduction of a purely {\it ad hoc} mass
scale which separates the perturbative from the nonperturbative QCD region \cite{margolis1,block7}. This mass
scale, as well as the fixed coupling constant present in the elementary cross sections, are adjusted in order to
obtain the best fits to the experimental data.

On the other hand, several recent works have shown that the gluon may develop a
dynamical mass (see the review \cite{alkofer} and the earlier work of Ref. \cite{cornwall}).
This dynamical gluon mass was already successfully introduced in the Pomeron model of Landshoff and
Nachtmann \cite{halzen}. Hence it is natural to ask if the arbitrary mass scale that appears in the
QCD-inspired models could be explained at a deeper level in terms of the dynamical gluon mass. This relation
seems to be a plausible possibility and in this paper we will show that the dynamical
gluon mass, as well as the IR finite coupling constant associated to it \cite{ans}, are in fact the
natural regulators for the cross sections calculations. Since the behavior of the running coupling constant
is constrained by the value of the dynamical gluon mass \cite{cornwall,ans}, we will be able to substitute
the two {\it ad hoc} parameters in the  ``mini-jet" models \cite{block,block2,block3}, namely
the infrared mass scale ($m_{0}$) and the effective value of the running coupling constant ($\alpha_{s}$), by
a physically well motivated one. In this way, beyond the natural interpretation of the arbitrary
infrared mass scale in terms of a dynamical one, it is possible to decrease the number of
parameters in this line of models.

The paper is organized as follows: In the next section we introduce a dynamical
gluon mass in the gluon-gluon scattering and compare the result to the standard one used in
QCD-inspired models. In Sec. III we develop a QCD-inspired eikonal model in the light of the calculation
of Sec. II. Our results are presented in the Sec. IV, where the best value for the dynamical gluon
mass is determined and used thereafter to determine several quantities of $pp$ and $\bar{p}p$ scattering. In
Sec. V we present our conclusions. 

\section{Infrared mass scale and gluon-gluon elastic scattering}

In recent years there have been discussions in the literature about how to merge in a doubtless
way the nonperturbative
QCD results with the perturbative expansion. It is worth mentioning that Brodsky has several times
called attention about the possibility to build up a skeleton expansion  where the nonperturbative
information would be included in vertices and propagators. In particular, the freezing of the
QCD running coupling constant at low energy scales could allow to capture at an inclusive level the
nonperturbative effects in a reliable way (see, for instance, Ref. \cite{brodsky}). The freezing of the
coupling constant and the existence of a dynamical gluon mass are intimately connected \cite{ans},
therefore they should appear systematically in this sought expansion.

It is possible that such
skeleton expansion could appear with the use of the pinch technique \cite{papa}. With this technique the
nonperturbative behavior of ``gauge invariant" propagators and vertices could be computed
nonperturbatively at one given
order and substituted into the perturbative skeleton expansion. The fact that the ``pinch'' parts help to
form gauge invariant quantities would result in well behaved matrix elements for the desired expansion.

It is clear that we are still far from the kind of expansion discussed above, and we have to rely on more
phenomenological approaches to go forward in this direction. One attempt to understand the effect
of dynamically massive gluons was performed by Forshaw, Papavassiliou and Parrinello \cite{papa2}, where
they do introduce bare massive gluons and study the amplitude behavior for some
tree and one-loop level diagrams that could be relevant for diffractive scattering. In this approach, for
example, the amplitude for the tree level process $q \bar{q} \rightarrow gg$ comes out with a  mass
dependence which is washed out in the high energy limit, but the massless limit is not recovered due to
the presence of a numerically small mass independent term. The calculation is instructive but does not
reproduce the high energy limit of massless gluons with two degrees of freedom, as discussed by Slavnov
many years ago \cite{slavnov}. Actually the dynamical
masses go to zero at large momenta and we should expect to recover the elementary cross
sections of perturbative QCD  in the high energy limit. Following this thought we could say that the sum
over the polarizations should be performed as if the gluons were massless otherwise we would not map
(at high energies) the desired skeleton expansion into the perturbative QCD expansion.

According to the above discussion, we cannot
work with a massive Yang-Mills theory or use a massive model where the third polarization state is
provided by a massless scalar field. Thus, we will just assume a phenomenological procedure
stated many years ago by Pagels and Stokar and named dynamical perturbation theory (DPT) \cite{pagels}.
The DPT approximation can be described as follows: amplitudes that do not vanish to all orders of
perturbation theory are given by their free-field values. On the other hand, amplitudes that vanish in all orders
in perturbation theory as $\propto \exp{(-1/g^2)}$ ($g$ is the coupling constant) are retained at lowest
order. In our case this means that the effects of the dynamical gluon mass in the propagators and vertices
will be retained, and the sum of polarizations will be performed for massless (free-field)
gluons, because its signal (for massive gluons) will not vanish from the elementary cross section. In this
approach, the differential elastic cross section for the process $gg\rightarrow gg$ is written as
\begin{eqnarray}
\frac{d\hat{\sigma}^{D\!PT}}{d\hat{t}}(\hat{s},\hat{t}\, ) = \frac{9\pi
\bar{\alpha}_{s}^{2}}{2\hat{s}^{2}} \left[ 3 
-\frac{\hat{s}[4M_{g}^{2} -\hat{s} - \hat{t}\, ]}{[\hat{t}-M_{g}^{2}]^{2}} 
-\frac{\hat{s}\hat{t}}{[3M_{g}^{2}-\hat{s}-\hat{t}\, ]^{2}}
-\frac{\hat{t}[4M_{g}^{2} - \hat{s} -\hat{t}\, ]}{[\hat{s}-M_{g}^{2}]^{2}} \right] ,
\label{h1}
\end{eqnarray}
where $\bar{\alpha}_{s}$ and $M_{g}^{2}$ are the expressions for the nonperturbative
running coupling constant and for the dynamical gluon mass, respectively. They were obtained by Cornwall
\cite{cornwall} by means of the pinch technique in order to derive a gauge invariant Schwinger-Dyson
equation for the gluon propagator. These expressions are given by
\begin{eqnarray}
\bar{\alpha}_{s} (q^{2})= \frac{4\pi}{\beta_0 \ln\left[
(q^{2}+4M_g^2(q^{2}))/\Lambda^2 \right]},
\label{acor}
\end{eqnarray}
\begin{eqnarray}
M^2_g(q^{2}) = m_g^2 \left[ \frac{\ln \left( \frac{q^{2}+4{m_g}^2}{\Lambda^2} \right) }{\ln
\left( \frac{4m_g^2}{\Lambda^2} \right) } \right]^{- 12/11} ,
\label{mdyna}
\end{eqnarray}
where $\beta_0 = 11- \frac{2}{3}n_f$ ($n_f$ is the number of flavors) and $\Lambda$($\equiv\Lambda_{QCD}$)
is the QCD scale parameter. The latter expression has been determined as a fit to the numerical solution for
the gluonic Schwinger-Dyson equation in the case of pure gauge QCD \cite{cornwall}. We also assume that the
introduction of fermions does not change drastically this behavior. The gluon mass scale $m_{g}$ has to be
found phenomenologically, and a typical value is $m_{g}=500 \pm 200$ MeV (for $\Lambda=300$ MeV)
\cite{cornwall, aguilar}. Note that we present the full $gg\to gg$ cross section just for
completeness, where the terms suppressed by powers of $\hat{s}$ are not important compared to leading
$\ln \hat{s}$ perturbative corrections. However it should be pointed out that up to now higher order
corrections have not been introduced in these type of models, and this is even more complicated
if we consider the nonperturbative effects that we are introducing in this work.

A different expression for the dynamical gluon mass can be found in Ref. \cite{aguilar3},
given by
\begin{eqnarray}
M^2_g(q^{2}) = \frac{m_{g}^{4}}{q^{2}+{m_g}^2} ,
\end{eqnarray}
which is consistent with the asymptotic behavior of $M_{g}(q^{2})$ in the presence of the gluon condensates
\cite{aguilar2}. However, the calculation of the hadronic cross section does not depend strongly on the
specific form of $M_{g}(q^{2})$, but more on its IR value (i.e., the value of $m_{g}$).

In the limit $q^{2}\gg \Lambda^{2}$, the dynamical mass $M_{g}(q^{2})$ vanishes, and
the nonperturbative QCD running coupling $\bar{\alpha}_{s}$ matches with the one-loop perturbative
QCD one. Thus, in the limit of large enough $q^{2}$, the expression (\ref{h1}) reproduces its
perturbative QCD counterpart:
\begin{eqnarray}
\frac{d\hat{\sigma}^{QCD}}{d\hat{t}}(\hat{s},\hat{t}\, ) = \frac{9\pi \alpha_{s}^{2}}{2\hat{s}^{2}}
\left[ 3 
+\frac{\hat{s}( \hat{s} + \hat{t}\, )}{\hat{t}^{2}} 
-\frac{\hat{s}\hat{t}}{(\hat{s}+\hat{t}\, )^{2}}
+\frac{\hat{t}(\hat{s} + \hat{t}\, )}{\hat{s}^{2}} \right] .
\label{h2}
\end{eqnarray}


To compute (\ref{h1}) we have used a vertex having a momentum dependent running coupling and a massive
gluon propagator in the Feynman gauge, where the sum over gluon polarizations was performed for massless
gluons.

The total cross section
$\hat{\sigma}(\hat{s})=\int_{\hat{t}_{min}}^{\hat{t}_{max}} (d\hat{\sigma}/d\hat{t}\, ) \, d\hat{t}$
for the subprocess $gg\rightarrow gg$, that will be used in the next section to compose the eikonal term
$\chi_{gg}$, is obtained by integrating over $4m_g^2 -\hat{s} \le \hat{t} \le 0$. In setting these kinematical
limits we have neglected the momentum behavior in Eq. (\ref{mdyna}), as expected from our discussion on
the weak dependence of hadronic cross sections on the specific form of $M_{g}^{2}(q^{2})$. A straightforward
calculation yields
\begin{eqnarray}
\hat{\sigma}^{D\!PT}(\hat{s}) &=& \frac{3\pi \bar{\alpha}_{s}^{2}}{\hat{s}} \left[ \frac{12\hat{s}^{4}
- 55 m_{g}^{2} \hat{s}^{3} + 12 m_{g}^{4} \hat{s}^{2} + 66 m_{g}^{6} \hat{s} -
8 m_{g}^{8}}{4 m_{g}^{2} \hat{s} [\hat{s} - m_{g}^{2}]^{2}} -
3 \ln \left( \frac{\hat{s} - 3m_{g}^{2}}{m_{g}^{2}}\right) \right] .
\label{h6}
\end{eqnarray}

The asymptotic energy ($\hat{s}$) dependence of the total cross section
$\hat{\sigma}^{D\!PT}(\hat{s})$ is of the following form
\begin{eqnarray}
\hat{\sigma}^{D\!PT}(\hat{s}) \approx \frac{9\pi \bar{\alpha}_{s}^{2}}{m_{g}^{2}} .
\label{h7}
\end{eqnarray}

We notice that the above result is similar to the asymptotic expression
for the gluon-gluon total elastic cross section usually adopted in QCD-inspired models (QIM),
\begin{eqnarray}
\hat{\sigma}^{QIM}(\hat{s}) \equiv \Sigma_{gg} = \frac{9\pi \alpha_{s}^{2}}{m_{0}^{2}},
\label{h8}
\end{eqnarray}
where the parameters $m_{0}$ and $\alpha_{s}$ are assumed to be equal to $0.6$ GeV and $0.5$,
respectively \cite{block2,block3}. We particularly call attention to these values, because they are of
the same order of magnitude
as the dynamical gluon mass scale ($m_g$) and its frozen IR value of the coupling constant, obtained
in other calculations of strongly interacting processes \cite{aguilar}. Therefore, all the
point in here is how to connect these nonperturbative results to the straightforward perturbative QCD
calculations.


\section{dynamical gluon mass and QCD-inspired eikonal models}

A consistent calculation of high-energy hadron-hadron cross sections must be compatible with analyticity
and unitarity constraints. The latter can be automatically satisfied by use of an eikonalized treatment
of the semihard parton processes. In an eikonal representation, the total, elastic and inelastic
cross sections are given by
\begin{eqnarray}
\sigma_{tot}(s) = 4\pi \int_{_{0}}^{^{\infty}} \!\! b\, db\, [1-e^{-\chi_{_{I}}(b,s)}\cos \chi_{_{R}}(b,s)],
\label{degt1}
\end{eqnarray}
\begin{eqnarray}
\sigma_{el}(s) = 2\pi \int_{_{0}}^{^{\infty}} \!\! b\, db\, |1-e^{-\chi_{_{I}}(b,s)+i\chi_{_{R}}(b,s)}|^2,
\label{degt2}
\end{eqnarray}
\begin{eqnarray}
\sigma_{in}(s) = \sigma_{tot}(s)- \sigma_{el}(s) =  2\pi \int_{_{0}}^{^{\infty}} \!\! b\, db\,
[1-e^{-2\chi_{_{I}}(b,s)}],
\label{degt3}
\end{eqnarray}
respectively, where $s$ is the square of the total center-of-mass energy and $\chi (b,s)$ is a complex
eikonal function: $\chi(b,s)=\chi_{_{R}}(b,s)+i\chi_{_{I}}(b,s)$. In this formalism, the factor
$e^{-2\chi_{_{I}}(b,s)}$ in the expression (\ref{degt3}) is interpreted as the probability that neither
nucleon is broken up in a collision at impact parameter $b$. The ratio $\rho$ of the real to the imaginary
part of the forward scattering amplitude is given by
\begin{eqnarray}
\rho (s) = \frac{\textnormal{Re} \{ i \int b\, db\, [1-e^{i\chi (b,s)}]  \}}{\textnormal{Im} \{ i \int b\,
db\, [1-e^{i\chi (b,s)}]  \}},
\label{degthyj1}
\end{eqnarray}
whereas the nuclear slope $B$ and the differential elastic scattering cross section are given by
\begin{eqnarray}
B(s) = \frac{\int b^{3}\, db\, [1-e^{i\chi (b,s)}]}{\int b\,
db\, [1-e^{i\chi (b,s)}]} ,
\label{degthyj2}
\end{eqnarray}
and
\begin{eqnarray}
\frac{d\sigma_{el}}{dt}(s,t)=\frac{1}{2\pi}\, \left| \int b\, db\, [1-e^{i\chi (b,s)}]\, J_{0}(qb) \right|^2 ,
\label{degthyj3}
\end{eqnarray}
respectively, where $J_{0}(x)$ is the Bessel function of the first kind. The eikonal
function can be written as a combination of an even and odd eikonal terms related by crossing symmetry. In
terms of the proton-proton ($pp$) and antiproton-proton
($\bar{p}p$) scatterings, this combination reads
$\chi_{pp}^{\bar{p}p}(b,s) = \chi^{+} (b,s) \pm \chi^{-} (b,s)$. 

Following the work of Block {\it et al.}
\cite{block3}, we write the even eikonal as the sum of  gluon-gluon, quark-gluon, and quark-quark
contributions:
\begin{eqnarray}
\chi^{+}(b,s) &=& \chi_{qq} (b,s) +\chi_{qg} (b,s) + \chi_{gg} (b,s) \nonumber \\
&=& i[\sigma_{qq}(s) W(b;\mu_{qq}) + \sigma_{qg}(s) W(b;\mu_{qg})+ \sigma_{gg}(s) W(b;\mu_{gg})] .
\label{final4}
\end{eqnarray}

Here $W(b;\mu)$ is the overlap function at impact parameter space and $\sigma_{ij}(s)$ is the elementary
subprocess cross section of colliding quarks and gluons ($i,j=q,g$). The overlap function is usually
associated with the Fourier transform of a dipole form factor,
\begin{eqnarray}
W(b;\mu) = \frac{\mu^2}{96\pi}\, (\mu b)^3 \, K_{3}(\mu b),
\end{eqnarray}
where $K_{3}(x)$ is the modified Bessel function of second kind. The $W(b;\mu)$ function is normalized
so that $\int d^{2}\vec{b}\, W(b;\mu)=1$. The odd eikonal $\chi^{-}(b,s)$, that accounts for the difference
between $pp$ and $\bar{p}p$ channels, is parametrized as
\begin{eqnarray}
\chi^{-} (b,s) = C_{odd} \, \Sigma_{gg} \, \frac{m_{0}}{\sqrt{s}} \, e^{i\pi /4}\, 
W(b;\mu_{odd}),
\end{eqnarray}
where $\Sigma_{gg}$ is given by the expression (\ref{h8}) and $m_{0}$ is an arbitrary IR mass scale.
$C_{odd}$ and $\mu_{odd}$ are fitting parameters. We borrow this term, with its correct analyticity
property, and write our odd eikonal as
\begin{eqnarray}
\chi^{-} (b,s) = C^{-}\, \Sigma \, \frac{m_{g}}{\sqrt{s}} \, e^{i\pi /4}\, 
W(b;\mu^{-}),
\label{bl2}
\end{eqnarray}
where $m_{g}$ is the dynamical gluon mass and the parameters $C^{-}$ and $\mu^{-}$ are constants to be
fitted. The factor $\Sigma$ is defined as
\begin{eqnarray}
\Sigma = \frac{9\pi \bar{\alpha}_{s}^{2}(0)}{m_{g}^{2}},
\end{eqnarray}
which is just the expression (\ref{h7}) deprived of any momentum dependence, with the coupling constant
$\bar{\alpha}_{s}$ set at its frozen IR value. This definition of
$\Sigma$, when compared with the $\Sigma_{gg}$ one, reveals explicitly the natural
relation between the infrared mass scales $m_{0}$ and $m_{g}$.

In the original Block {\it et al.} model the eikonal functions $\chi_{qq} (b,s)$ and
$\chi_{qg} (b,s)$, needed to
describe the lower-energy forward data, are parametrized with terms dictated by the Regge
phenomenology. Similarly, we parametrize our quark-quark and quark-gluon contributions as
\begin{eqnarray}
\chi_{qq}(b,s) = i \, \Sigma \, A \,
\frac{m_{g}}{\sqrt{s}} \, W(b;\mu_{qq}),
\label{mdg1}
\end{eqnarray}
\begin{eqnarray}
\chi_{qg}(b,s) = i \, \Sigma \left[ A^{\prime} + B^{\prime} \ln \left( \frac{s}{m_{g}^{2}} \right) \right] \,
W(b;\sqrt{\mu_{qq}\mu_{gg}}),
\label{mdg2}
\end{eqnarray}
where $A$, $A^{\prime}$, $B^{\prime}$, $\mu_{qq}$ and $\mu_{gg}$ are fitting parameters. Notice that in
the above expression the inverse size (in impact parameter) $\mu_{qg}$ is defined in the same way as in the
original model, i.e., $\mu_{qg}\equiv \sqrt{\mu_{qq}\mu_{gg}}$. In deriving the expressions (\ref{mdg1})
and (\ref{mdg2}) we have used the fact that the main contribution to the asymptotic behavior of hadron-hadron
total cross sections comes from gluon-gluon semihard collisions, since $g(x)\gg q(x)$ at small-$x$ values.
Therefore it is enough to build instrumental quark-quark and quark-gluon parametrizations for the
expected high-energy behavior of the $pp$ and $\bar{p}p$ amplitudes and to compute only the gluon-gluon contribution
by means of the calculation procedure described in the last section. For example, the term $\ln (s/m_g^2)$
is naturally explained by the presence of a massive gluon in the $qg\to qg$ subprocess. In this way the chosen
eikonals reflect exactly the terms that come from such cross section calculations. We also have not
considered the effect of dynamically generated quark masses in the subprocesses involving quarks. This
approach involves an extra parameter ($m_{q} \approx 250-300$ MeV), but we believe that its effect is smaller
compared to the dynamical gluon mass one.

The gluon-gluon contribution dominates at high energy and determines the asymptotic behavior of the total
cross section. In our model we associate the gluon eikonal term $\chi_{gg}(b,s)$ (see the expression (\ref{final4}))
with the cross section $\sigma_{gg}^{D\!PT}(s)$: $\chi_{gg}(b,s)\equiv \sigma_{gg}^{D\!PT}(s)W(b; \mu_{gg})$.
Hence the gluon eikonal contribution includes $gg\to gg$ subprocesses with colour nonsinglet exchange in all
possible channels. The
cross section $\sigma_{gg}^{D\!PT}(s)$ is written as
\begin{eqnarray}
\sigma_{gg}^{D\!PT}(s) = C^{\prime} \int_{4m_{g}^{2}/s}^{1} d\tau \,F_{gg}(\tau)\,
\hat{\sigma}^{D\!PT} (\hat{s}) ,
\label{sloh1}
\end{eqnarray}
where $F_{gg}(\tau)$ is the convoluted structure function for pair $gg$, $\hat{\sigma}^{D\!PT}(\hat{s})$ is
the subprocess cross section given by expression (\ref{h6}), and $C^{\prime}$ is a fitting parameter. In the
above expression we have introduced the energy threshold $\hat{s}\geq 4m_{g}^{2}$ for the final state gluons,
assuming that these are screened gluons, in a procedure similar to the calculation of Ref. \cite{cornwall2}. The
structure function $F_{gg}(\tau)$ is written as
\begin{eqnarray}
F_{gg}(\tau)=[g\otimes g](\tau)=\int_{\tau}^{1} \frac{dx}{x}\, g(x)\,
g\left( \frac{\tau}{x}\right),
\end{eqnarray}
where $g(x)$ is the gluon distribution function, usually adopted as
\begin{eqnarray}
g(x) = N_{g} \, \frac{(1-x)^5}{x^{J}},
\label{distgf}
\end{eqnarray}
where $J=1+\epsilon$ and $N_{g}=\frac{1}{240}(6-\epsilon)(5-\epsilon)...(1-\epsilon)$. In this definition the
term $\sim 1/x^{1+\epsilon}$ simulates the effect of scaling violations in the small $x$ behavior of $g(x)$
\cite{block7}. In the Regge language the quantity $J$, that controls the asymptotic behavior of
$\sigma_{tot}(s)$, is the so called intercept of the Pomeron. In fact, neglecting the variation
with $q^{2}$ of the asymptotic expression (\ref{h7}), it is possible to show that
\begin{eqnarray}
\lim_{\hat{s}\to \infty} \int^{1}_{4m_{g}^{2}/s} d\tau \, F_{gg}(\tau)\, \hat{\sigma}^{D\!PT} (\hat{s})
\sim \left( \frac{s}{4m_{g}^{2}} \right)^{\epsilon} .
\label{gtso2}
\end{eqnarray}

Hence the total cross section behaves asymptotically as a Pomeron power law $s^{J-1}$, and a consistent
value of $J$ can be determined by fitting forward quantities data through a Regge pole model. Recently, by
means of an extended Regge model, some authors have determined the bounds for the soft Pomeron intercept
imposed by the accelerator and cosmic ray data currently available \cite{luna1,luna2}. These results are
consistent with a Pomeron intercept $J=1.085$ (specifically, $J-1 = 0.085\pm0.006$ in the case of constrained
bounds \cite{luna2}), and corroborate the choice of the Pomeron intercept value adopted is this work.

We ensure the correct analyticity properties of our model amplitudes by substituting $s\to se^{-i\pi/2}$
throughout Eqs. (\ref{mdg1}), (\ref{mdg2}) and (\ref{sloh1}). For simplicity, we will refer to our
QCD-inspired model with a dynamical gluon mass simply as the DGM model.

\section{results}

In all the fits performed in this paper we use a $\chi^{2}$ fitting procedure,
where the value of $\chi^{2}_{min}$ is distributed as a $\chi^{2}$ distribution with N degrees of
freedom (DOF). The fits to the experimental data sets are performed adopting an interval
$\chi^{2}-\chi^{2}_{min}$ corresponding, in the case of normal errors, to the projection of the
$\chi^{2}$ hypersurface containing 90\% of probability. In the case of the DGM model (8 fitting parameters)
this corresponds to the interval $\chi^{2}-\chi^{2}_{min}=13.36$ \cite{num}. To determine the optimum value
for the dynamical gluon mass and extract the best phenomenological values of the DGM model
parameters, we follow a two step process. First, we select specific input values for the dynamical gluon
mass and carry out global
fits to all high-energy forward $pp$ and $\bar{p}p$ scattering data above $\sqrt{s}=10$ GeV and to the
elastic differential scattering cross section for $\bar{p}p$ at $\sqrt{s}=1.8$ TeV. These forward data sets
include the total cross section ($\sigma_{tot}$), the ratio of the real to imaginary part of the forward
scattering amplitude ($\rho$), and the nuclear slope in the forward direction ($B$). 
We use the data sets compiled and analyzed by the Particle Data Group \cite{hagiwara}, to which we
add the new E811 data on $\sigma^{\bar{p}p}_{tot}$ and $\rho^{\bar{p}p}$ at $\sqrt{s}=1.8$
TeV \cite{avila1}. The statistic and systematic errors of the forward quantities have been added in
quadrature. The input values of
the $m_{g}$ have been chosen to lie in the interval $[300, 800]$ MeV, as suggested by the
value $m_{g}=500\pm200$ usually obtained in other calculations of strongly interacting processes (see section
II). Although no physical argument ensures that the optimum value of $m_{g}$ lies in the chosen input
mass interval, our global fit results indicate a minimum value just about $m_{g}\approx 400$ MeV.
These results are shown in Fig. 1, where a general dashed curve is added to guide the eye. Roughly, taking a
5\% variation on the minimal $\chi^{2}/DOF$ value indicated by the general curve, it is possible to estimate a
dynamical gluon mass $m_{g}\approx 400^{+350}_{-100}$ MeV. This result is totally compatible with the ones of
Ref. \cite{halzen}: $m_{g}= 370$ MeV.

Next, in order to determine the parameters of the DGM model, we set the value of the dynamical gluon mass
to $m_{g}=400$ MeV (optimal value) and carry out a global fit only to all high-energy forward
$pp$ and $\bar{p}p$ scattering data above $\sqrt{s}=10$ GeV, not including the elastic differential
scattering cross section $d\sigma^{\bar{p}p} /dt$ at $\sqrt{s}=1.8$ TeV. The values of the fitted parameters
are given in Table 1. The $\chi^{2}/DOF$ for this global fit was 1.075 for 188 degrees of freedom. The results
of the fits to $\sigma_{tot}$, $\rho$ and $B$ for both $pp$ and $\bar{p}p$ channels are displayed in Figs. 2,
3 and 4, respectively, together with the experimental data. Within this procedure, the Tevatron
differential cross section, as well as the Tevatron-run II and the CERN LHC ones, can be predicted by
the DGM model. These predictions are shown in Fig. 5. Table II contains predictions for the forward
quantities at these energies, where the quoted errors are the statistical errors due to the errors in
the fitted parameters.

\section{Conclusions}

In this paper we have investigated the influence of an infrared dynamical gluon mass scale in the calculation
of $pp$ and $\bar{p}p$ forward scattering quantities through a QCD-inspired eikonal model. By means of the
dynamical perturbation theory (DPT), we have computed the tree level $gg\to gg$ cross section taking into
account the dynamical gluon mass, and have shown that the IR divergences associated with the
gluon-gluon subprocess cross section are naturally regulated by this dynamical scale. In order to make a
connection between the total subprocess cross section $\hat{\sigma}_{gg}(\hat{s})$ and the forward
$pp$ and $\bar{p}p$ quantities, we have developed a QCD-inspired eikonal model where the onset of the
dominance of gluons in the interaction of high energy hadrons is managed by the dynamical gluon mass scale.
Using this formalism it was possible not only to reduce the number of parameters of the model, but also
to give a consistent physical explanation for each one. For example, in some recent papers on QCD-inspired
models \cite{block2,block3}, the two arbitrary constants $m_0$ and $\alpha_s$ were assumed to be equal $0.6$ GeV and
$0.5$, respectively; in our approach the IR value of running coupling constant is driven by the dynamical gluon
mass, i.e., its IR behavior depends on the value of
$m_{g}$. This connection permit us to decrease the number of parameters required to describe the hadronic
experimental data.

By means of a global fit to the forward $pp$ and $\bar{p}p$ scattering data and to $d\sigma^{\bar{p}p} /dt$
data at $\sqrt{s}=1.8$ TeV, we have determined the best phenomenological value of the dynamical gluon mass,
namely $m_{g}\approx 400^{+350}_{-100}$ MeV. Interestingly enough, this value
is of the same order of magnitude as
the value $m_{g}\approx 500 \pm 200$ MeV, obtained in other calculations of strongly interacting processes.
This result corroborates theoretical analysis taking into account
the possibility of dynamical mass generation and show that, in principle, a dynamical nonperturbative
gluon propagator may be used in
calculations as if it were a usual (derived from Feynman rules) gluon propagator. 

With the dynamical gluon mass set at $m_{g}=400$ MeV, we have performed a global fit only to the forward
$pp$ and $\bar{p}p$ scattering data, in the same way as is usually performed in the former QCD-inspired
models. Our model allows us to describe successfully the forward scattering quantities
$\sigma_{tot}$, $\rho$ and $B$, as well as to predict the $\bar{p}p$ differential cross section at
$\sqrt{s}=1.8$ TeV in excellent agreement with the available experimental data. These results show
that the DGM model is well suited for detailed predictions of the forward quantities to be measured at higher
energies. In particular, for the total cross sections to be measured at Tevatron-run II and CERN-LHC energies,
the model predicts the values $\sigma_{tot}=75.7\pm5.4$ mb and $\sigma_{tot}=102.9\pm7.1$ mb, respectively.
Our central LHC value prediction is close to the central one in the Ref. \cite{block3}, namely
$\sigma_{tot}=108$ mb. This relatively small difference reflects the fact that the dynamical gluon
mass and its associated coupling constant successfully replace the values of the {\it ad hoc} parameters of the
former QCD inspired models. However, if the $pp$ total cross section is measured at the LHC with a precision up
to 5\%, a selection between these QCD models may be possible.

In summary, we argue that the QCD-inspired eikonal model with a dynamical IR mass scale provides an useful
phenomenological tool to the study of the hadron-hadron diffractive scattering, where a purely perturbative
QCD method is inadequate.

\begin{acknowledgments}
This research was supported by the Conselho Nacional de Desenvolvimento Cient\'{\i}fico e
Tecnol\'ogico-CNPq under contract 151360/2004-9 (EGSL, AAN), and by the Funda\c{c}\~ao de Amparo
\`a Pesquisa do Estado de S\~ao Paulo-FAPESP under constracts 2004/10619-9 (MJM) and 2003/00928-1 (AM).
\end{acknowledgments}

\newpage

\begin{table*}
\caption{Values of the parameters of the DGM model resulting from the global fit to the forward $pp$ and
$\bar{p}p$ data. The dynamical gluon mass scale was set to $m_{g}=400$ MeV.}
\begin{ruledtabular}
\begin{tabular}{cc}
$C^{\prime}$ & (12.097$\pm$0.962)$\times 10^{-3}$ \\
$\mu_{gg}$ [GeV]& 0.7242$\pm$0.0172 \\
$A$ & 6.72$\pm$0.92 \\
$\mu_{qq}$ [GeV] & 1.0745$\pm$0.0405 \\
$A^{\prime}$ & (4.491$\pm$0.179)$\times 10^{-3}$ \\
$B^{\prime}$ & 1.08$\pm$0.14 \\
$C^{-}$ & 3.17$\pm$0.35 \\
$\mu^{-}$ [GeV]& 0.6092$\pm$0.0884 \\
\end{tabular}
\end{ruledtabular}
\end{table*}

\begin{table*}
\caption{Predictions of the $pp$ and $\bar{p}p$ forward scattering quantities $\sigma_{tot}$,
$\rho$ and $B$ for the Fermilab Tevatron run-II (TEVII) and the CERN LHC energies.}
\begin{ruledtabular}
\begin{tabular}{ccc}
& TEVII [1.96 TeV]& LHC [14 TeV]\\
\hline
$\sigma^{pp}_{tot}$, $\sigma^{\bar{p}p}_{tot}$ [mb]& 75.7$\pm$5.4 & 102.9$\pm$7.1\\
$\rho^{pp}$, $\rho^{\bar{p}p}$ & 0.129$\pm$0.009 & 0.114$\pm$0.005\\
$B^{pp}$, $B^{\bar{p}p}$ [GeV$^{-2}$] & 16.97$\pm$0.99 & 19.36$\pm$1.12\\
\end{tabular}
\end{ruledtabular}
\end{table*}

\begin{figure}
\label{difdad}
\vspace{2.0cm}
\begin{center}
\includegraphics[height=.60\textheight]{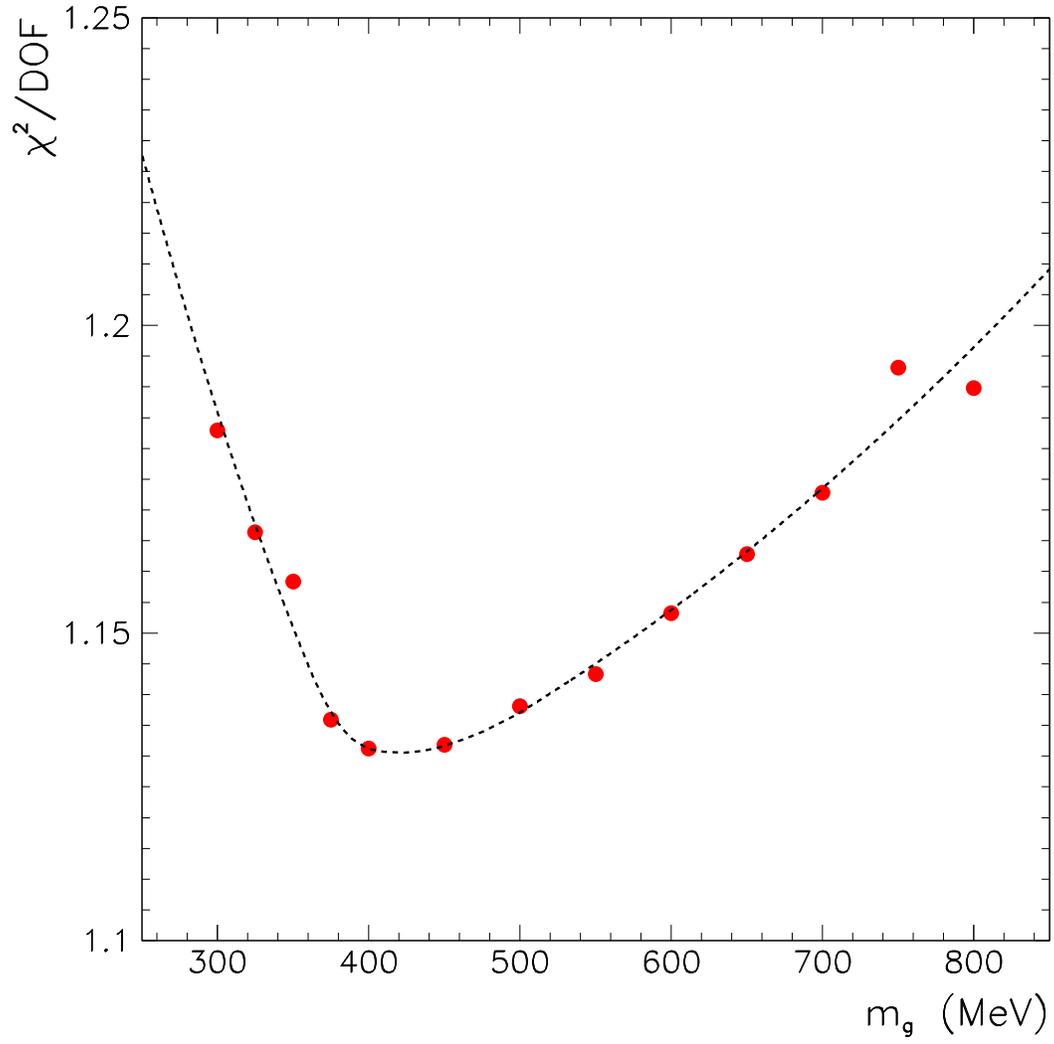}
\caption{The $\chi^{2}/DOF$ as a function of dynamical gluon mass $m_{g}$. }
\end{center}
\end{figure}

\begin{figure}
\label{difdad}
\vspace{2.0cm}
\begin{center}
\includegraphics[height=.60\textheight]{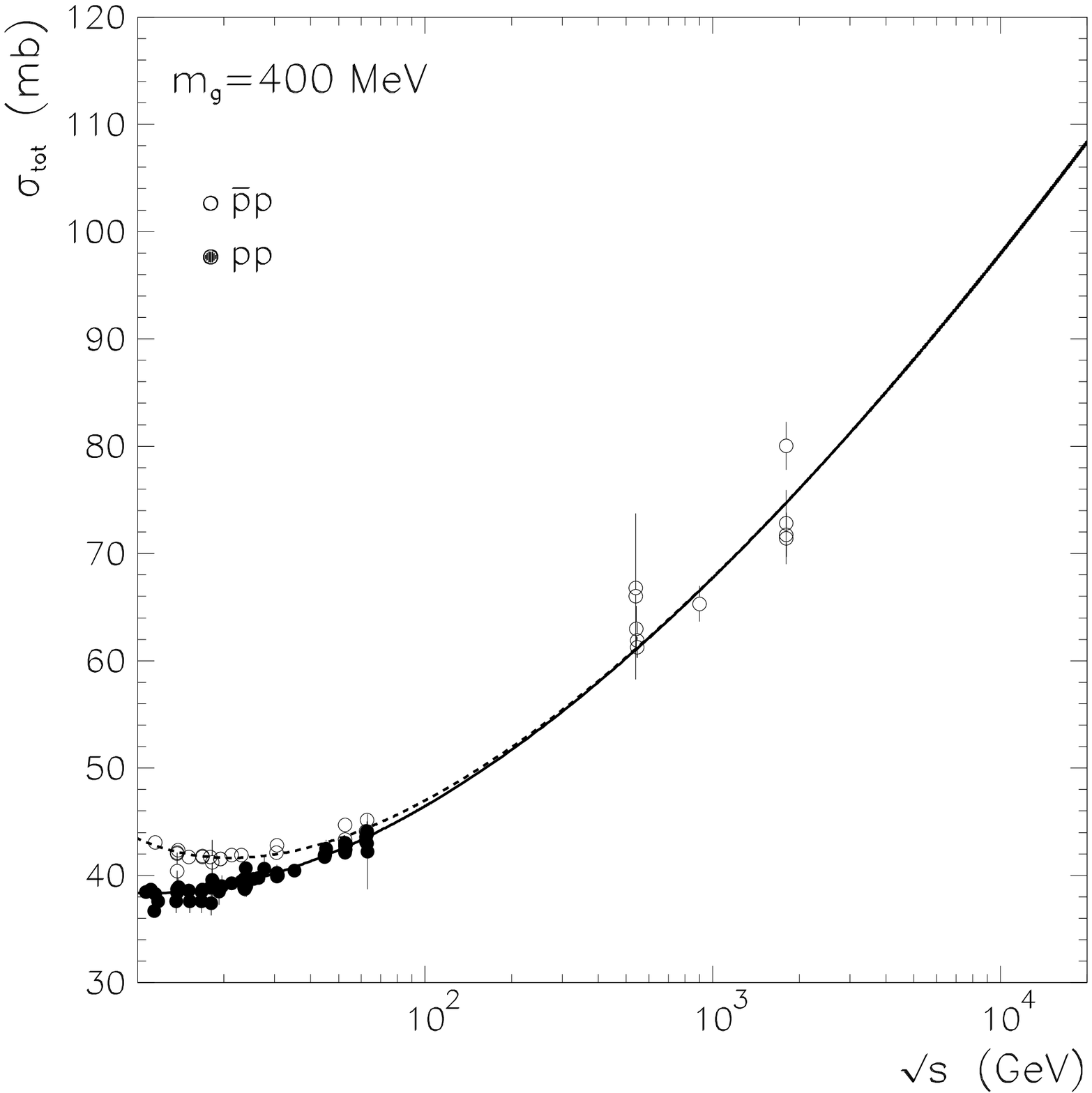}
\caption{Total cross section for $pp$ (solid curve) and $\bar{p}p$ (dashed curve) scattering.}
\end{center}
\end{figure}

\begin{figure}
\label{difdad}
\vspace{2.0cm}
\begin{center}
\includegraphics[height=.60\textheight]{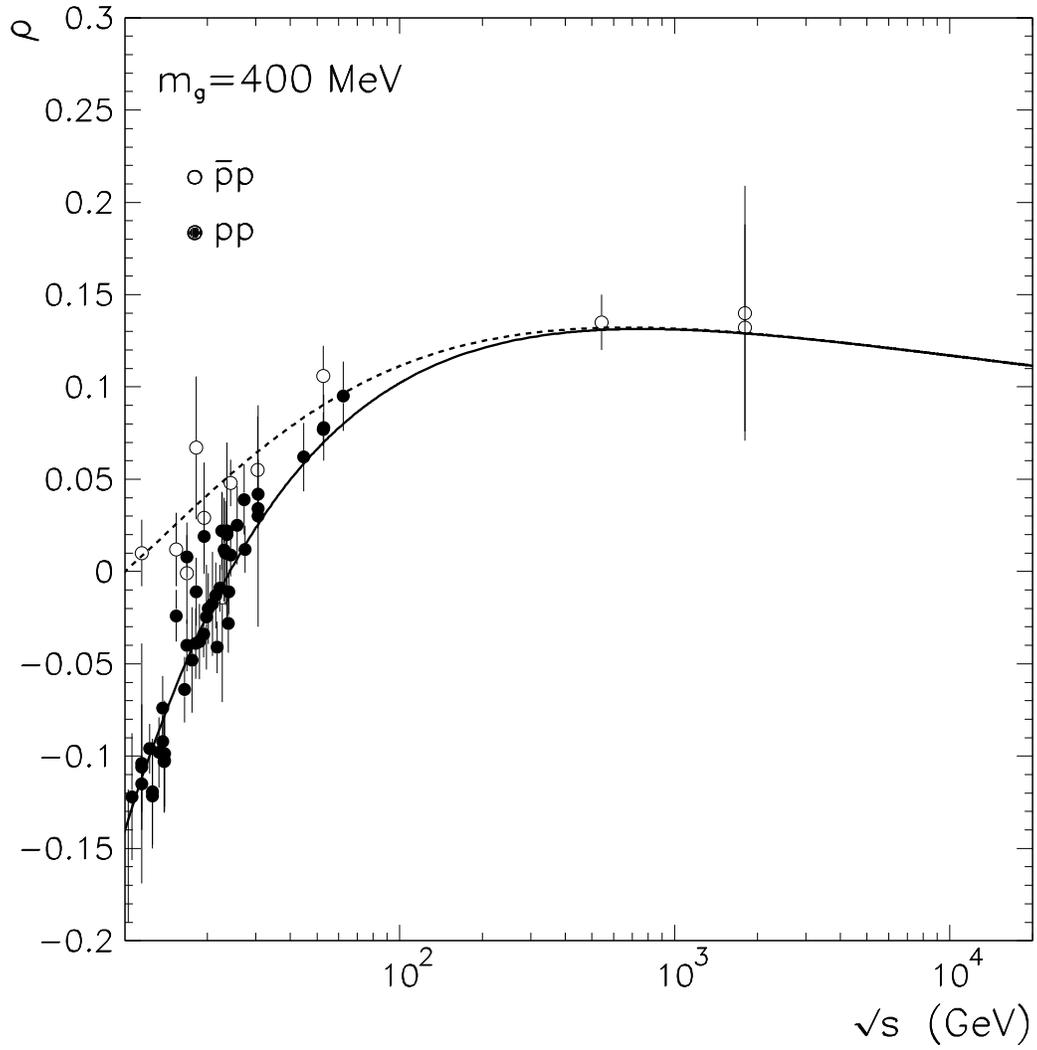}
\caption{Ratio of the real to imaginary part of the forward scattering amplitude for $pp$ (solid curve)
and $\bar{p}p$ (dashed curve) scattering. }
\end{center}
\end{figure}

\begin{figure}
\label{difdad}
\vspace{2.0cm}
\begin{center}
\includegraphics[height=.60\textheight]{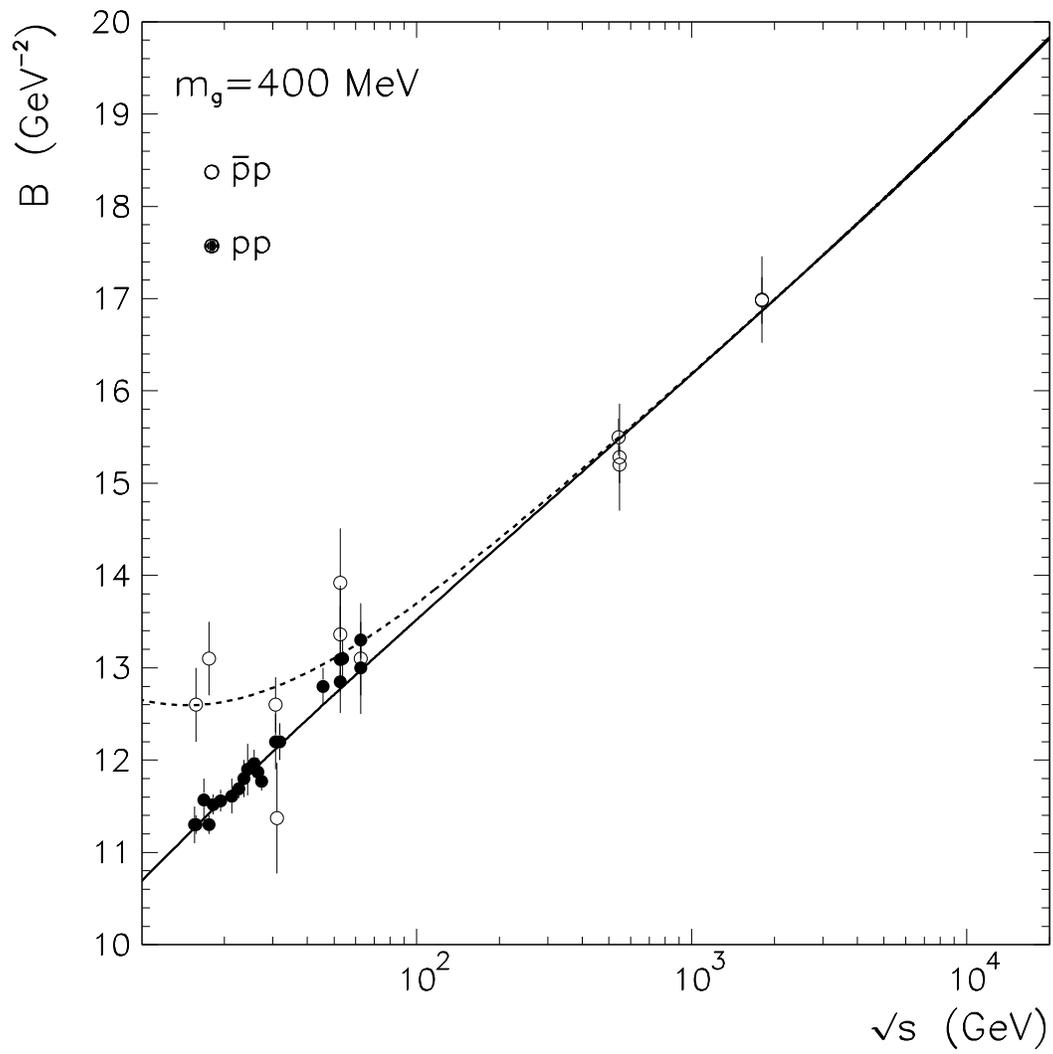}
\caption{Nuclear slope parameter for elastic $pp$ (solid curve) and $\bar{p}p$ (dashed curve) scattering. }
\end{center}
\end{figure}

\begin{figure}
\label{difdad}
\vspace{2.0cm}
\begin{center}
\includegraphics[height=.60\textheight]{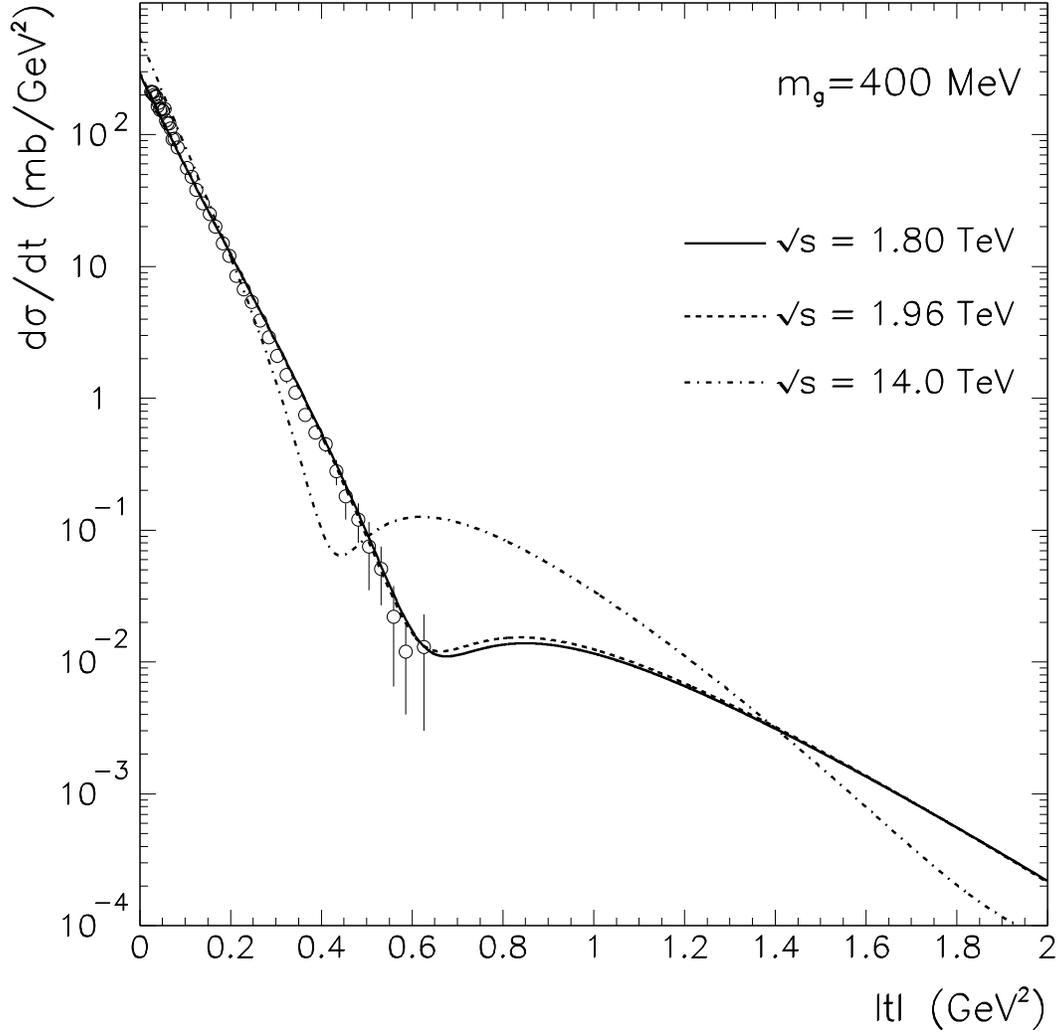}
\caption{Predictions for the elastic differential scattering cross sections at
$\sqrt{s}=1.8, 1.96$ and $14$ TeV. In our model the channels $pp$ and $\bar{p}p$ are not distinguished
at high energies. The data points are from E710 \cite{E710}.}
\end{center}
\end{figure}

\end{document}